# Singlet Oxygen Generation as a Major Cause for Parasitic Reactions during Cycling of Aprotic Lithium-Oxygen Batteries

Nika Mahne[1], Bettina Schafzahl[1], Christian Leypold[1], Mario Leypold[3], Sandra Grumm[1], Anita Leitgeb[1], Gernot A. Strohmeier[3,4], Martin Wilkening[1], Olivier Fontaine[5,6], Denis Kramer[7], Christian Slugovc[1], Sergey M. Borisov[2] and Stefan A. Freunberger[1]*

[1] Institute for Chemistry and Technology of Materials, Graz University of Technology, Stremayrgasse 9, 8010 Graz, Austria

[2] Institute for Analytical Chemistry and Food Chemistry, Graz University of Technology, Stremayrgasse 9, 8010 Graz, Austria

[3] Institute of Organic Chemistry, Graz University of Technology, Stremayrgasse 9, 8010 Graz, Austria

[4] Austrian Centre of Industrial Biotechnology (acib) GmbH, Petersgasse 14, 8010 Graz, Austria

[5] Institut Charles Gerhardt Montpellier, UMR 5253, CC 1701, Université Montpellier, Place Eugène Bataillon, 34095 Montpellier Cedex 5, France

[6] Réseau sur le Stockage Electrochimique de l'Énergie (RS2E), FR CNRS

[7] Engineering Sciences, University of Southampton, SO17 1BJ, Southampton, UK

* to whom correspondence should be addressed: freunberger@tugraz.at

**Non-aqueous metal-oxygen batteries depend critically on the reversible formation/decomposition of metal oxides on cycling. Irreversible parasitic reactions cause poor rechargeability, efficiency, and cycle life and have predominantly been ascribed to the reactivity of reduced oxygen species with cell components. These species, however, cannot fully explain the side reactions. Here we show that singlet oxygen forms at the cathode of a lithium-oxygen cell during discharge and from the onset of charge, and accounts for the majority of parasitic reaction products. The amount increases during discharge, early stages of charge, and charging at higher voltages, and is enhanced by the presence of trace water. Superoxide and peroxide appear to be involved in singlet oxygen generation. Singlet oxygen traps and quenchers can reduce parasitic reactions effectively. Awareness of the highly reactive singlet oxygen in non-aqueous metal-oxygen batteries gives a rationale for future research towards achieving highly reversible cell operation.**

Rechargeable non-aqueous metal-$O_2$ (air) batteries have attracted immense interest because of their high theoretical specific energy and potentially better sustainability and cost in comparison to current lithium-ion batteries[1-5]. Cell chemistries include Li-, Na- and K-$O_2$ with the Li-$O_2$ cell being most intensely studied[6-9]. Charge is stored at the cathode by the reversible formation/decomposition of metal oxides on discharge/charge[10,11]. In the Li-$O_2$ cell this is typically $Li_2O_2$. Practical realization, however, still faces many challenges[5,8,12-14]. Perhaps the most significant obstacle arises from severe parasitic reactions during cycling[3-5,7,8,10,11,13-26]. These reactions decompose the electrolyte as well as the porous electrode (typically carbon with binder) and cause poor rechargeability, high charging voltages, low efficiency, buildup of parasitic reaction products, and early cell death within a few cycles.

Many researchers have investigated the origin of parasitic reactions and proposed strategies to mitigate them[7,8,10,16-19]. Superoxide has been most widely mentioned in causing side reactions on discharge since it forms as an intermediate in $O_2$ reduction and is a strong nucleophile and base[3,11,14,20,21,27]. Also, $Li_2O_2$ was found to react with the electrolyte and carbon on discharge[3,21-24]. These reactivities were used to explain the observation that on discharge typically close to the ideal value of 2 electrons per one $O_2$ molecule are consumed despite significant amounts of side products such as $Li_2CO_3$, Li formate and Li acetate being formed[17,24]. On charge typically the $e^-/O_2$ ratio deviates significantly from 2 and more of the side products form[5,7,15,24,25]. These parasitic reactions occur at charging potentials well within the stability window (oxidative stability) of carbon and electrolyte in the absence of $Li_2O_2$[21,23,25]. It was therefore suggested that some sort of reactive intermediates of $Li_2O_2$ oxidation cause electrolyte and carbon decomposition on charge[11,23,25,28].

Chemical oxidation of alkaline peroxides in non-aqueous media is known to generate singlet oxygen ($^1\Delta_g$ or $^1O_2$), the first excited state of triplet ground state dioxygen ($^3\Sigma_g^-$)[29-32]. Based on the reversible potential of $Li_2O_2$ formation and the energy difference between triplet and singlet oxygen, the formation of $^1O_2$ in the Li-$O_2$ cell has been hypothesized to be possible at charging potentials exceeding 3.5 to 3.9 V vs. Li/Li$^+$[11,23]. Only recently $^1O_2$ was reported to form in small quantities between 3.55 and 3.75 V[28]. Overall, the hitherto known processes cannot consistently explain the observed irreversibilities. Only better knowledge of parasitic reactions may allow them to be inhibited so that progress towards fully reversible cell operation can continue.

Here we show that $^1O_2$ forms in the Li-$O_2$ cathode during discharge and from the onset of charge and that it is responsible for a major fraction of the side products in the investigated system with ether electrolyte. The lower abundance on discharge and higher one on charge can consistently explain the typically observed deviations of the $e^-/O_2$ ratio from the ideal value of 2. The origin of the $^1O_2$ on charge appears to be superoxide and peroxide. The presence of trace water enhances the formation during both discharge and charge. We also show that $^1O_2$ traps and quenchers as electrolyte additives can significantly reduce the amount of side products associated with $^1O_2$.

**Reactivity of the electrolyte with singlet oxygen**

The discharge product formed at the Li-$O_2$ cathode in relatively stable electrolytes, such as the widely used glyme (oligo-ethylene glycol dimethyl ether) based ones, consists predominantly of $Li_2O_2$ accompanied by a typical pattern of side products including $Li_2CO_3$, Li acetate and Li formate[12,15,21,23,24,26,33]. The same side products form upon oxidation of $Li_2O_2$ (charging) and eventually release $CO_2$ and other fragments at sufficiently oxidizing potentials[21,25]. A large body of work has identified the reduced $O_2$ species superoxide and peroxide or their lithium compounds to trigger the formation of these products[3,11,14,15,20,21,34]. To investigate whether $^1O_2$ would lead to the same products, we generated it inside a typical electrolyte, 0.1 M lithium perchlorate ($LiClO_4$) in ethylene glycol dimethyl ether (DME), and analysed the formed products, Fig. 1. $^1O_2$ was generated photochemically by illuminating the $O_2$ saturated electrolyte containing a small concentration of a photosensitizer inside a closed vessel (for experimental details see the Methods section). The head space was then purged to a mass spectrometer (MS) for analysis to detect readily evolved gases, and, after addition of acid, to detect whether $Li_2CO_3$ had formed, Fig. 1a. MS analysis does not show any detectable direct $CO_2$ evolution, but $CO_2$ evolving from $Li_2CO_3$. A second portion of equally treated electrolyte was dissolved in $D_2O$ and subjected to $^1H$-NMR spectroscopy, Fig. 1b. The $^1H$-NMR spectrum confirms the presence of Li formate and Li acetate *via* the peaks for HCOOD and $CH_3COOD$ that form upon contact with $D_2O$. The literature on the reactivity of $^1O_2$ with organic substrates most commonly states peroxides as an initial product[32]. Going along this line, we assume that $^1O_2$ produces ROOH, R•, and ROO• as the first reactive intermediates of electrolyte degradation more efficiently than the reduced oxygen species, which were proposed to initiate electrolyte degradation via the same intermediates albeit high activation energies have been noted[33,35,36]. Taken together, these results show that the typical pattern of parasitic reaction products formed during discharge and charge of Li-$O_2$ cells could to some extent originate from the presence of $^1O_2$.

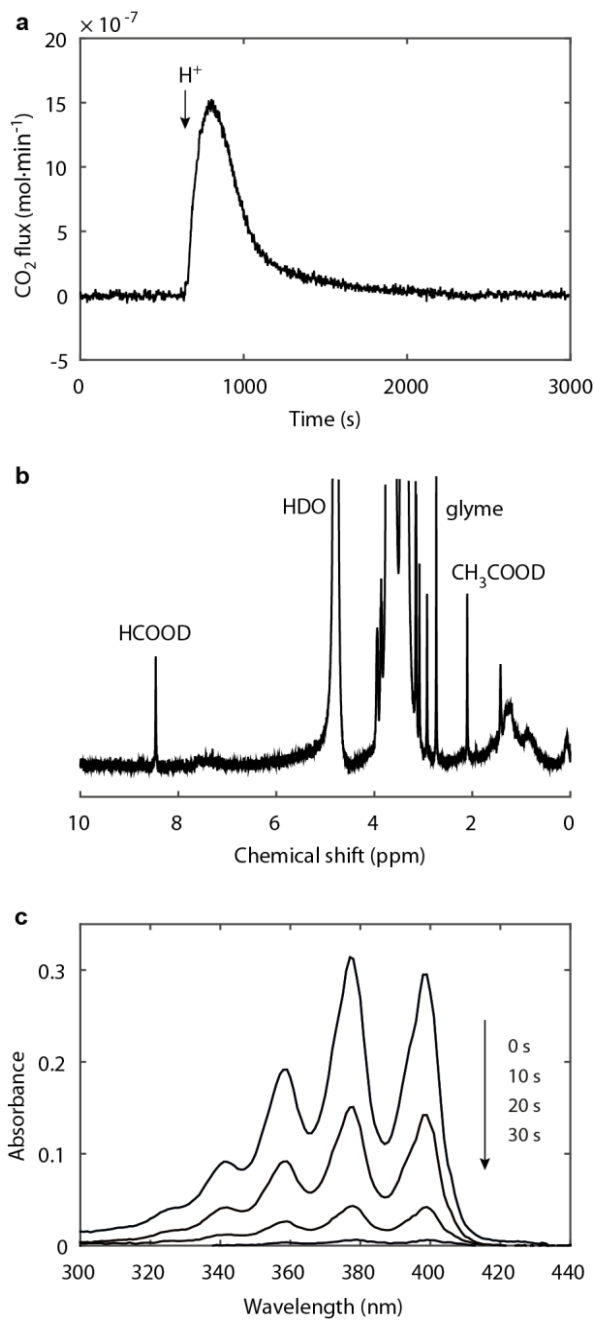

**Figure 1 | Reactivity of the electrolyte with singlet oxygen. a,** $CO_2$ evolution measured by mass spectrometry above $O_2$ saturated 0.1 M $LiClO_4$ in DME electrolyte that has been exposed for 30 min to $^1O_2$. The latter has been produced *in situ* by photogeneration with the sensitizer palladium(II) *meso*-tetra(4-fluorophenyl)tetrabenzoporphyrin with the head space closed. After illumination the head space was purged to the mass spectrometer and $H_3PO_4$ added at the time indicated to evolve $CO_2$ from $Li_2CO_3$. **b,** $^1$H-NMR spectrum of the equally treated electrolyte when dissolved in $D_2O$. **c,** UV-Vis absorption spectra as a function of illumination time of the same electrolyte that additionally contained $2.6×10^{-5}$ M 9,10-dimethylanthracene. The absorbance in the ordinate of **c** is dimensionless, thus there is no unit.

*Operando* detection of singlet oxygen in the Li-O$_2$ cathode

Probing whether $^1O_2$ is involved in the cell reaction requires sensitive methods that are compatible with the cell environment. So far described methods for detection of $^1O_2$ are either based on direct detection of characteristic light emissions upon decay into the ground state, or on the selective reactivity with probe molecules that are themselves interrogated by spectroscopic means[32]. The short lifetime of $^1O_2$ in liquid media, various competing decay routes, and low sensitivity of NIR detectors make the detection of the specific emission of $^1O_2$ at 1270 nm challenging and insensitive[31,32,37]. Therefore, the absence of a detectable signal would provide no definite proof for the absence of $^1O_2$. Nevertheless, we could detect this emission for a case with high $^1O_2$ abundance as discussed later.

To detect $^1O_2$ at quantities, which would be responsible even for small amounts of parasitic products, we devised a sensitive and selective method with a chemical probe compatible with the cell environment at any stage of cycling. Previously, highly sensitive probes for aqueous media have been described, which contain a quencher group attached to a chromophore and show fluorescence "switch on" upon reaction of the quenching group with singlet oxygen[32,38]. However, the chromophores used so far are not electrochemically inert in the relevant potential range of ~2 to 4 V vs. Li/Li$^+$. Typically used chromophores include fluorescein and rhodamine, which all undergo electrochemical reactions in this range, Supplementary Fig. 1. The quencher group for $^1O_2$ is typically a substituted anthracene derivative such as 9,10-dimethylanthracene (DMA) or 9,10-diphenylanthracene (DPA), which form the corresponding endoperoxide. DPA itself has been used directly as $^1O_2$ probe based on the decrease of absorbance as a sign for the presence of $^1O_2$ [39]. We have added 2.6×10$^{-5}$ M DMA to 0.1 M LiClO$_4$ in DME and exposed the solution to *in situ* photogenerated $^1O_2$; all DMA was consumed within less than a minute, which indicates rapid reaction with $^1O_2$ to its endoperoxide (DMA-O$_2$) in this environment, Fig. 1c. DPA reacted in the same experiment approximately two orders of magnitude slower, Supplementary Fig. 2. Cyclic voltammograms with 2 mM DMA and 0.1 M LiClO$_4$ in DME under Ar show electrochemical stability between 1.8 and 4.5 V, Supplementary Fig. 3. The DMA was then transformed to its endoperoxide (DMA-O$_2$) by means of *in situ* photogenerated $^1O_2$. Cyclic voltammograms taken thereafter show likewise stability of the DMA-O$_2$ between 2.5 and >4.5 V. Inertness against superoxide ($O_2^-$) and hence selectivity for $^1O_2$ was confirmed by stirring DMA with an excess of $KO_2$ in DME containing 0.1 M $LiClO_4$ and taking $^1$H-NMR spectra and UV-Vis spectra at time intervals up to 22 h. The results do not show any detectable decomposition products of the DMA, Supplementary Fig. 4 to 7. Taken together, the above experiments confirm DMA to be a sensitive and selective probe for $^1O_2$ in the cell environment. We use DMA in the following first as a fluorescent probe for *operando* detection of $^1O_2$. *Operando* fluorescence requires a relatively low DMA concentration in the μM range, which somewhat restricts the detection limit. Later we use DMA in the mM range to detect $^1O_2$ with maximum sensitivity and to remove it, which requires measuring the conversion of DMA to DMA-$O_2$ by *ex situ* HPLC.

An *operando* fluorescence setup as detailed in the Methods section was constructed. Briefly, the cell consisted of a porous carbon working electrode in an $O_2$ saturated electrolyte containing $1.6 \times 10^{-5}$ M DMA and 0.1 M $LiClO_4$ in tetraethylene glycol dimethyl ether (TEGDME). As the counter electrode we used $Li_{1-x}FePO_4$ to exclude reactivity of the DMA with a Li metal anode. The cell was assembled inside a gas tight quartz cuvette with slightly pressurized $O_2$ head space. Excitation and emission wavelengths were chosen according to the respective maxima in these spectra of DMA, Supplementary Fig. 8. The electrolyte was stirred to ensure constant $O_2$ concentration irrespective of consumption or evolution during cycling. This is important since $O_2$ is a fluorescence quencher and changes in intensity could otherwise stem from changing $O_2$ concentration as shown in Supplementary Fig. 9.

Results for charging a cathode, containing chemically produced $Li_2O_2$ with 0.1 M $LiClO_4$ in TEGDME as the electrolyte, by applying voltage steps are shown in Fig. 2a. At voltages up to 3.5 V the fluorescence intensity, reflecting DMA concentration, remains unchanged within the measurement accuracy. Above this voltage, the signal drops with increasing rate. The cumulatively consumed DMA corresponds to ~2% of the theoretically evolved $O_2$ (based on charge) being $^1O_2$. This value is the lower boundary of the actual abundance since at the low DMA concentration (~0.5% of the $O_2$ concentration) competing sinks for $^1O_2$ other than reaction with DMA can be expected to dominate.

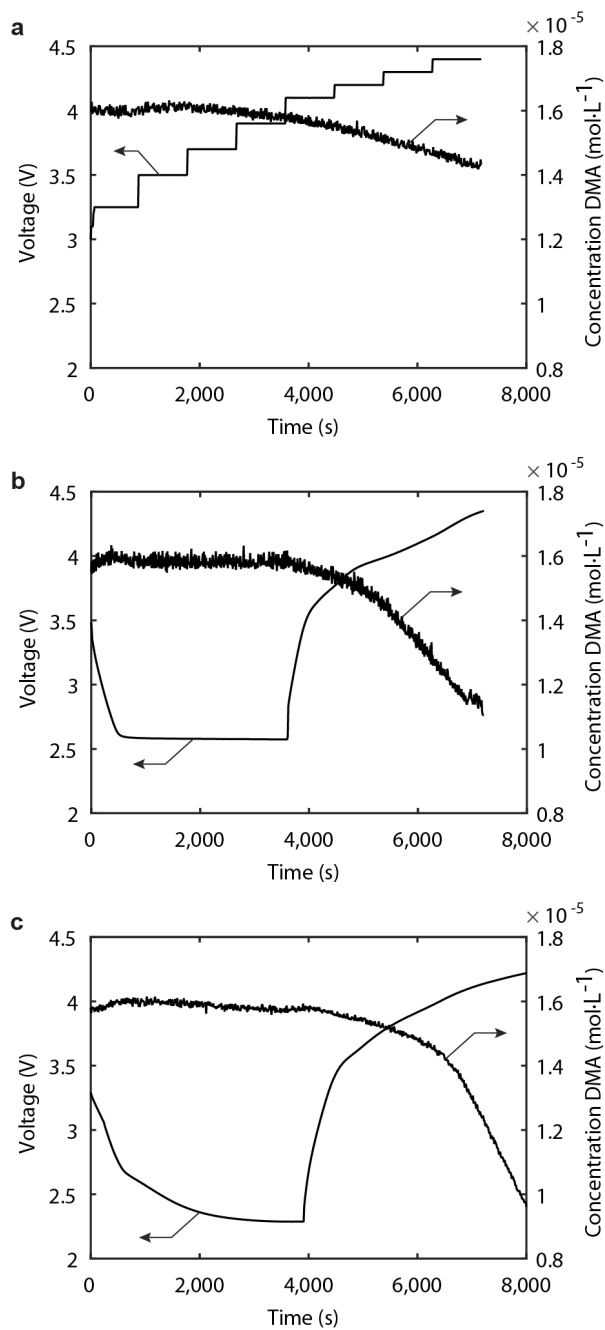

**Figure 2 |** *Operando* **fluorescence spectroscopy during Li-O₂ cell operation with electrolytes containing 9,10-dimethylanthracene (DMA) as singlet oxygen trap. a**, Potentiostatic oxidation of a carbon black electrode containing chemically produced $Li_2O_2$ in $O_2$ saturated 0.1 M $LiClO_4$ in TEGDME. Voltage steps and DMA concentration. **b**, Galvanostatic discharge and charge of a carbon black electrode at 25 μA·cm$^{-2}$ in dry $O_2$ saturated 0.1 M $LiClO_4$ in TEGDME. Voltage profile and DMA concentration. **c**, Galvanostatic discharge and charge of a carbon black electrode at 25 μA·cm$^{-2}$ in $O_2$ saturated 0.1 M $LiClO_4$ in TEGDME containing 1000 ppm water. Voltage profile and DMA concentration. All electrolytes had an initial DMA concentration of 1.6×10$^{-5}$ M.

To probe whether $^1O_2$ is also formed during discharge, we cycled electrodes in the fluorescence setup. Results for galvanostatic cycling of a porous carbon cathode in dry 0.1 M $LiClO_4$ in TEGDME are shown in Fig. 2b. Upon discharge the DMA concentration remains nearly unchanged within the measurement accuracy. However, immediately after switching to charging, starting from ~3 V, the signal drops with increasing slope as charging progresses and the voltage climbs towards 4.3 V, where full recharge is reached. The cumulatively consumed DMA corresponds to ~4% of the expected $O_2$ being $^1O_2$. The results demonstrate that $^1O_2$ was formed from the very beginning of charge at a significant rate. However, this experiment could not tell with certainty whether $^1O_2$ was formed on discharge. A possible source of $^1O_2$ on discharge is the reaction of the superoxide intermediate, the first step of $O_2$ reduction, with trace water, which has been shown to result in $^1O_2$[37]. Therefore, we have run an analogous experiment with 1000 ppm $H_2O$ in the electrolyte, Fig. 2c. DMA consumption was seen throughout discharge at an approximately constant rate. Again, the rate of DMA consumption increased substantially immediately after the cell was switched to charging and increased as charge progressed to higher voltages. The consumption during charging is increased in comparison to the dry electrolyte and reaches a value of ~6% of the expected $O_2$ being $^1O_2$. This suggests that trace water contributes to the formation of $^1O_2$ on discharge and charge. Since the required DMA concentration for *operando* fluorescence is low, the abundance of $^1O_2$ on discharge appears to be close to the detection limit. Clear evidence for $^1O_2$ on discharge comes from *ex situ* measurements with ~2000 times the DMA concentration as discussed later.

We have shown above that DMA is very reactive with $^1O_2$ and reactive to a negligible extent with superoxide in the cell environment. To unambiguously show that there is indeed $^1O_2$ formation and that the DMA consumption does not originate from possible other reactive oxygen species, we measured the specific emission of $^1O_2$ at 1270 nm (for experimental details see the Methods section). This radiative decay to the ground state gives a very weak signal, which is further attenuated by competing sinks for $^1O_2$. One such is deactivation with the solvent with a strongly solvent dependent lifetime[31]. Attempts to detect radiative decay in the ether electrolyte proved fruitless, which is explicable by the short lifetime in this solvent. Therefore, we have chosen deuterated acetonitrile, where the $^1O_2$ lifetime is higher than in ethers and also higher than in the non-deuterated solvents[32]. We also added 1000 ppm $D_2O$ since above experiments have shown higher $^1O_2$ generation when trace water was present, besides that the lifetime is longer in $D_2O$ than in $H_2O$. Results for galvanostatic cycling are shown in Fig. 3. The signal/noise ratio does not permit a clear statement about the abundance of $^1O_2$ during discharge, which is to be expected given the low generation rate detected with *operando* fluorescence in Fig. 2. In accord with above results there is, however, unambiguous proof of $^1O_2$ generation from the start of charging and increasing rate as charging progresses to higher voltages.

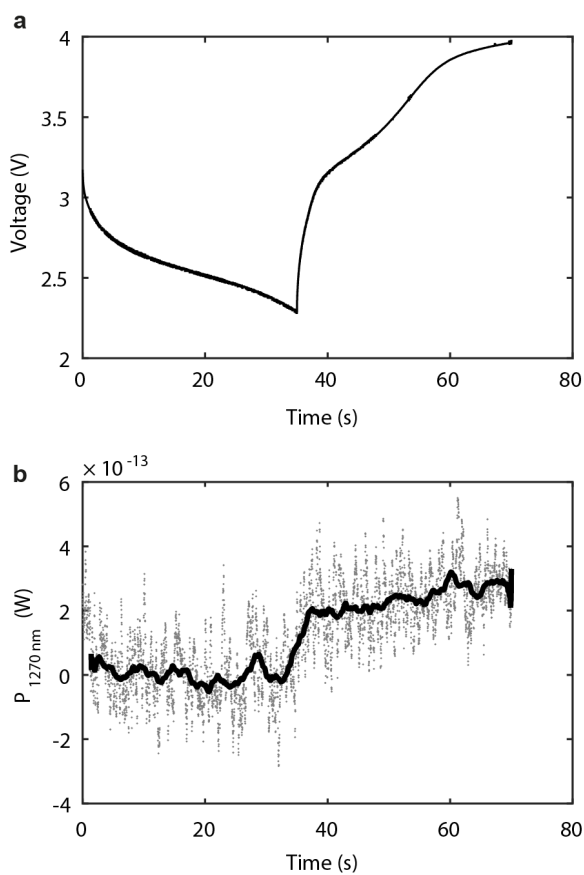

**Figure 3 |** *Operando* **NIR emission measurement during cycling of a Li-$O_2$ cathode. a**, Voltage profile during galvanostatic reduction and oxidation of an Au-grid electrode at 0.12 mA·cm$^{-2}$ in $O_2$ saturated 0.1 M LiClO$_4$ in deuterated acetonitrile containing 1000 ppm $D_2O$. **b**, The power of the optical emission at 1270 nm. The grey trace represents the sensor signal and the black trace the moving average to guide the eye.

**Trapping and quenching singlet oxygen**

The above results show that $^1O_2$ forms in significant quantities from the start of charging and suggest a smaller abundance during discharge. To estimate the fraction of the parasitic products during discharge and charge that originates from $^1O_2$ and to investigate whether removing the $^1O_2$ before it can react with cell components would effectively reduce these parasitic reactions, we examined the effect of $^1O_2$ trapping and quenching. The former removes $^1O_2$ in a chemical reaction and the latter deactivates it by physical quenching, *e.g., via* a temporary charge transfer complex[40]. Trapping is, however, irreversible and physical quenching is therefore preferred because neither quencher nor $O_2$ is consumed. The literature suggests a variety of quenchers including aliphatic amines and quinones[41]. We have chosen DMA as $^1O_2$ trap since it is effective in the cell environment, and 1,4-diazabicyclo[2.2.2]octane (DABCO) as quencher since it has been reported to be effective in non-aqueous environment[41]. DABCO also allows access to a relevant potential range between ~2.0 and 3.6 V and is stable with superoxide (Supplementary Fig. 10 and 11).

Li-$O_2$ cells with porous carbon black electrodes were constructed as described in the Methods section. Three electrolytes were used: 0.1 M LiClO$_4$ in TEGDME that either contained no additive, 30 mM DMA, or 10 mM DABCO. Cycling was carried out at constant current in $O_2$ atmosphere. Cells were cycled to various discharge and charge capacities, then stopped and subjected to further analysis. A typical load curve is shown in Fig. 4a. As DABCO is oxidized at ~3.6 V, cells containing this additive were only recharged to 3.5 V and then held there until the first recharge capacity was reached.

To quantify the amount of carbonaceous side products (Li$_2$CO$_3$ and Li carboxylates) formed at each stage of discharge and charge, the electrodes were analysed with a previously established procedure[25]. It involves treating the washed electrodes with acid to decompose the Li$_2$CO$_3$ present, followed by treatment with Fenton's reagent to oxidize the Li carboxylates. The evolved CO$_2$ was quantified by mass spectrometry and the results are presented in Fig. 4b. DMA as a $^1O_2$ trap is consumed and forms the corresponding endoperoxide DMA-O$_2$. We take advantage of this feature for quantifying the $^1O_2$ present in the cell by measuring the conversion of DMA to DMA-O$_2$ by means of HPLC, Fig. 4c and no additive, 30 mM DMA, or 10 mM DABCO. Cells were stopped and analyzed at the capacities indicated by the red squares. **b**, Amount of carbonaceous side reaction products per mg carbon. **c**, Fraction of the initial DMA that has reacted to DMA-$O_2$ *via* the reaction with $^1O_2$ in the cells that contained DMA as additive.

Supplementary Fig. 12. Here we use 30 mM DMA in the electrolyte, which is close to saturation and ~2000 times the concentration used in the fluorescence experiments. It can therefore be expected that a large fraction of any $^1O_2$ present, albeit not necessarily all, will react with the DMA instead of other cell components. At the same time, DMA becomes a significantly more sensitive probe for $^1O_2$ than as used for fluorescence.

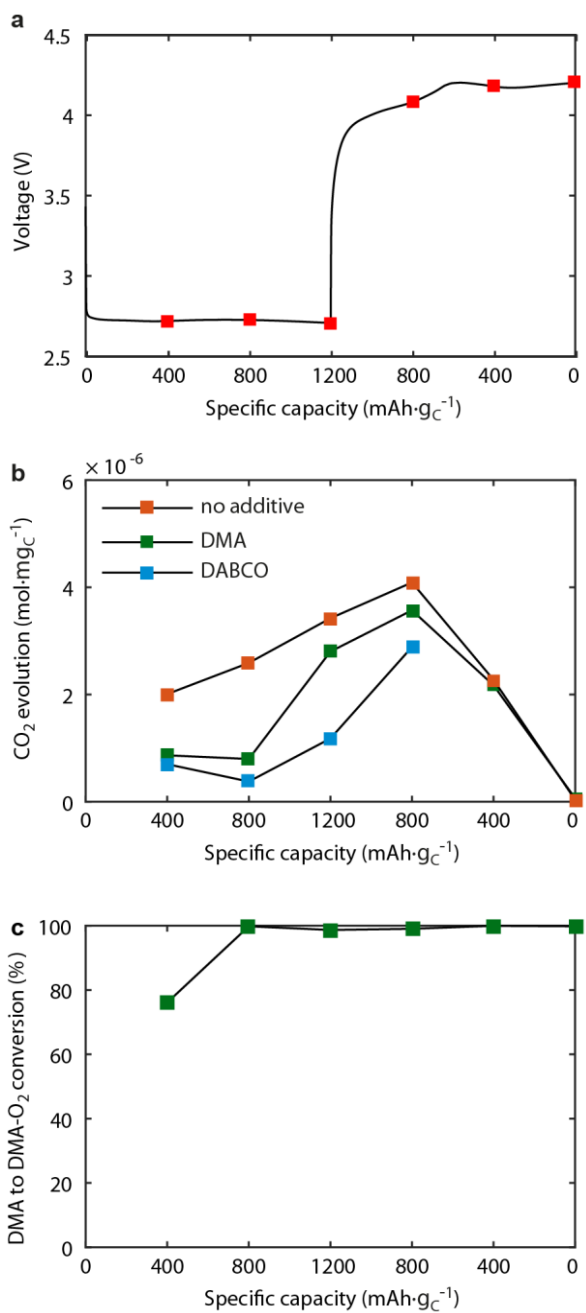

**Figure 4 |** *Ex situ* **analysis of Li-O$_2$ cathodes run with electrolytes without or with $^1O_2$ trap DMA or quencher DABCO. a**, Representative voltage profile during galvanostatic discharge and charge of a porous carbon black electrode at 70 mA·g$_C^{-1}$ in O$_2$ saturated 0.1 M LiClO$_4$ in TEGDME containing either

Considering the cell without additive, there is continuous growth of the amount of side products with increasing discharge capacity. The amount further increases to the sampling point at one third recharge, and then vanishes nearly completely towards full recharge. This is in accord with previous investigations on the build-up and removal of the side products during cycling[15,24,25]. It was shown that on discharge side products originate predominantly from the electrolyte. At early stages of charge, the electrolyte further decomposes to solid products, accompanied by $Li_2CO_3$ from the carbon electrode[24,25]. As charge continues to higher voltages, carbon decomposition becomes more significant and carbon and electrolyte decomposition go along with $CO_2$ evolution from already present parasitic products. So far it was reasoned that side products during discharge stem from the reactivity of cell components with the superoxide intermediate or $Li_2O_2$[3,11,14,20-24,27]. It is worth noting that carbon is considered stable on oxidation well beyond 4 V in the absence of $Li_2O_2$ and so is the ethereal electrolyte[15,21,25]. Carbon corrosion and electrolyte decomposition at lower charging voltages were therefore tentatively associated with intermediates of $Li_2O_2$ oxidation[23,25]. Our *operando* fluorescence results show that recharging the cell forms $^1O_2$ from the very start of charging and that it is responsible for at least part of the carbon and electrolyte decomposition from the start of charge.

Turning to the cells with $^1O_2$ trap or quencher, a significant reduction of side products during discharge is evident for both additives, Fig. 4b. Considering first the cell with DMA, the side products amount to between a half and a third of those without additive up to the second sampling point. Thereafter, the side products grow close to the level without the DMA. This can be explained considering the conversion of DMA to DMA-$O_2$, Fig. 4c. At the first sampling point, 76% of the initially present DMA was consumed, and it was fully consumed at the second point. Thereafter, no effect on side product formation can be expected as is seen in the carbonate/carboxylate data, Fig. 4b. By considering the charge passed at the first sampling point and the DMA conversion, a ratio of ~1 mol DMA consumed per 10 mol of $O_2$ reduced can be determined.

Turning to the cells with DABCO as quencher, side products amount to consistently less than in the case of DMA additive and to 10 to 30% of the additive-free case on discharge, Fig. 4b. From these values, we can estimate the fraction of parasitic products on discharge originating from $^1O_2$ to be at least 70%. DABCO is also effective upon charging and significantly reduces the side products at the first sampling point on charge. We assume the reason for the lower efficiency on charging to be the much higher $^1O_2$ generation on charge than on discharge as we have shown above with *operando* fluorescence. With DABCO we could, however, not recharge the cell fully due to the electrochemical stability limit of 3.5V, Supplementary Fig. 10. Quenchers need to be efficient in the cell environment, electrochemically stable and inert in contact with superoxide and $Li_2O_2$. These conditions are also the ones distinguishing quenchers required for the Li-$O_2$ cell from previous uses of quenchers[32].

To prove the effective $^1O_2$ removal by the trap over an entire cycle and to estimate the fraction of the parasitic products during charge that originates from $^1O_2$, we performed *operando* electrochemical mass spectrometry (OEMS) experiments with cells containing either no additive or 30 mM DMA. Figure 4c has shown that at a discharge capacity of 400 mA·$g_C^{-1}$ ~75% of the DMA had been converted to DMA-$O_2$. Therefore the mass spectrometry cells were discharged to only 200 mAh·$g_C^{-1}$ to ensure that most of the DMA was still present at the end of discharge and could act on charge. The results for charge are shown in the Fig. 5 and the full data in Supplementary Fig. 13 and 14. During discharge the $e^-/O_2$ ratio is close to the theoretical value of 2, with the ratio being closer with DMA (2.01 $e^-/O_2$) than without (2.11 $e^-/O_2$).

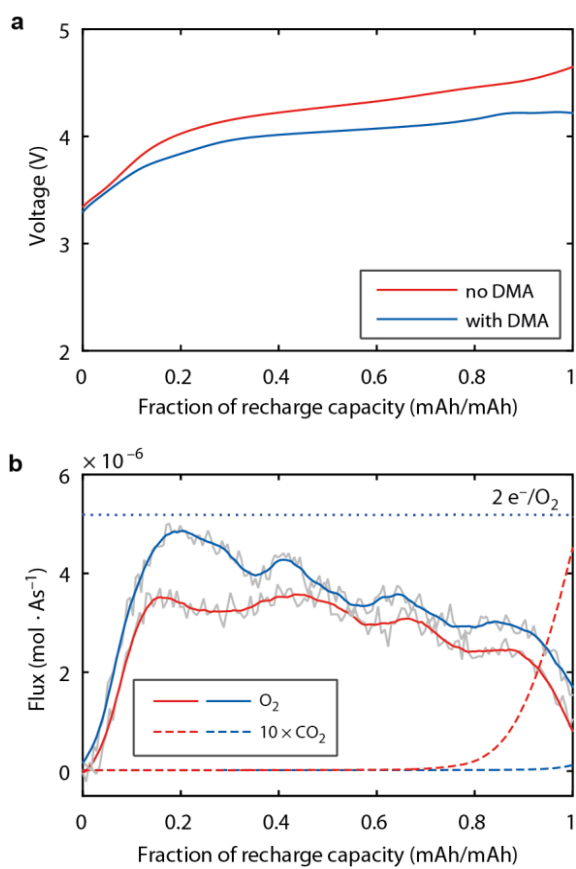

**Figure 5 |** *Operando* **electrochemical mass spectrometry of Li-$O_2$ cathodes run with electrolytes containing either no additive or the $^1O_2$ trap DMA. a, b,** Voltage profiles (**a**) and fluxes of $O_2$, $e^-$, and $CO_2$ (**b**) during galvanostatic charge after the cells have been discharged to 200 mAh·$g_C^{-1}$. The porous carbon black electrodes were run at 100 mA·$g_C^{-1}$ in 0.1 M $LiClO_4$ in TEGDME containing either no additive or 30 mM DMA. The grey traces in **b** represent the measurements and the blue and red traces the moving average to guide the eye.

With the DMA additive the recharge voltage is lower throughout than without DMA. With DMA the $O_2$ evolution reaches ~93% of the theoretical value at the beginning and fades to ~2/3 towards the end of charge. Without DMA the $O_2$ evolution is significantly lower throughout charging and reaches a maximum of 2/3 of the theoretical value. An even stronger difference is seen in the $CO_2$ evolution. Significant $CO_2$ evolution without DMA is contrasted by a 30 fold reduced $CO_2$ amount with DMA (based on the integral peak area in Supplementary Fig. 14b and f). The strong reduction of the $CO_2$ amount in combination with the observed $O_2$ evolution suggests that the majority of the parasitic products that form during charge at voltages below the oxidative stability limit of electrolyte and carbon are due to the occurrence of $^1O_2$. A more in depth discussion for this assignment is given in the section Supplementary Discussion in the Supplementary Information. Taken together, the trap and quencher experiments contribute more evidence that $^1O_2$ is responsible for the majority of side products upon discharge and charge, and that suitable additives can effectively reduce side reactions. The required oxidation stability of such additives can be reduced by using redox mediators that greatly reduce the charging voltage[8,42,43].

**Pathways to singlet oxygen**

The results are consistent with $^1O_2$ being to a large part responsible for commonly reported observations about the $O_2$ balance and side products. First, on discharge the $e^-/O_2$ ratio is typically found within several percent of the ideal value of 2 despite significant amounts of side products such as $Li_2CO_3$, Li formate and Li acetate with $Li_2O_2$ yields reported below 90%[24-26]. Second, on charge the $e^-/O_2$ ratio typically deviates significantly by more than 10% from the ideal value of 2 from the start with the deviation increasing as charging progresses[24]. This deviation goes along with the formation of more of the mentioned solid side products until the charging voltage is sufficiently high to oxidize them to release $CO_2$ and other fragments[15,25,26]. So far the formation of these products could not be consistently explained by the reactivity of the known reactive species superoxide and peroxide alone[14,23-25]. The previous hypothesis that $^1O_2$ can form by charging $Li_2O_2$ via $Li_2O_2 \rightarrow O_2 + 2Li^+ + 2e^-$ at voltages exceeding 3.5 to 3.9 V has only recently been verified with small quantities forming between 3.55 and 3.75 V[11,23,28]. The formation of $^1O_2$ on discharge and on charge below 3.5 V was, however, neither suggested on theoretical grounds nor shown experimentally before.

On discharge one possible source of $^1O_2$ is the disproportionation of $LiO_2$ according to

$$2\ LiO_2 \rightarrow (LiO_2)_2 \rightarrow Li_2O_2 + {}^1O_2 \qquad (1)$$

This pathway via a $(LiO_2)_2$ dimer appears plausible when the structures and energies of some of these dimers as calculated by Bryantsev et al. are considered[44]. They found low lying isomers in both the triplet and singlet state. They have given reaction free energies for the disproportionation reaction via the lowest triplet dimer to yield $^3O_2$ and it is reasonable to assume that the reaction proceeding via a singlet dimer will yield $^1O_2$. When $H_2O$ or other proton sources are available the superoxide will be protonated to form HOO• that has been reported to either undergo reduction by superoxide or disproportionate and to be able to release in either case $^1O_2$ in the overall reaction, which is more detailed in the Supplementary Discussion[37,45].

$$2\,O_2^- + 2H^+ \rightarrow H_2O_2 + {}^1O_2 \qquad (2)$$

Overall we propose the disproportionation of superoxide in the presence of either $Li^+$ or $H^+$ as the $^1O_2$ source on discharge. On charge we suggest three possible pathways. First, we suggest an analogous path to the one on discharge involving disproportionation of superoxide in the presence of either $Li^+$ or $H^+$. It has been suggested that the first step of charging $Li_2O_2$ involves a deintercalation at the surface to form $LiO_2$-like surface species ($Li_2O_2 \rightarrow LiO_2 + Li^+ + e^-$) that further disproportionate to evolve $O_2$ in an overall 2 $e^-/O_2$ process[46-48]. Here $^1O_2$ may form analogously to Eq. (1). Similarly, $^1O_2$ may form from proton sources (such as $H_2O$) reacting with the $LiO_2$-like surfaces according to Eq. (2). This pathway for $^1O_2$ formation can be active from the onset of charge as soon as $Li^+$ and $e^-$ are extracted. Second, a further 1 $e^-$ oxidation of the surface $LiO_2$ species ($LiO_2 \rightarrow O_2 + Li^+ + e^-$) could give $^1O_2$ in an overall 2 $e^-/O_2$ process. Thermodynamically, $^1O_2$ formation from electrochemical oxidation of superoxide is possible above $E^0_{O_2/LiO_2} + E(^1\Delta_g \leftarrow {}^3\Sigma_g^-)$. The thermodynamic equilibrium potential $E^0_{O_2/LiO_2}$ was estimated to be between 2.29 and 2.46 V [9,47,49]. With an energy difference of ~1 eV between $^1O_2$ and $^3O_2$, a thermodynamic voltage for $^1O_2$ evolution of 3.26 to 3.43 V can be estimated. Finally, above ~3.55 V the known pathway sets in as suggested before by Scrosati et al. and shown by Gasteiger et al. with $^1O_2$ evolving from electrochemical oxidation of $Li_2O_2$ in a 2 $e^-/O_2$ process ($Li_2O_2 \rightarrow O_2 + 2Li^+ + 2e^-$). Note that superoxide is both a proficient source and efficient quencher of $^1O_2$ via Eq. (3) [50].

$$O_2^- + {}^1O_2 \rightarrow {}^3O_2 + O_2^- \qquad (3)$$

We therefore believe that our observation of less $^1O_2$ on discharge and more on charge in the ether electrolyte results at least in part from the differing abundance of superoxide that can reduce the $^1O_2$ lifetime by quenching, which counteracts equally superoxide concentration driven formation. More precisely, net formation of $^1O_2$ will depend on the relative kinetics of all superoxide sources and sinks (with $^1O_2$ being involved in both) and not solely on the superoxide concentration. These sources and sinks are both electrochemical and chemical and change with discharge and charge, electrolyte, current, and potential. We thus further suggest that the current density and electrolyte properties will influence the $^1O_2$ formation in much the same way it governs the occurrence of superoxide on discharge and charge below 3.5V[5,33]. Further, charge current will drive $^1O_2$ production if it causes charging voltages above 3.5 V.

**Conclusions**

By combining complementary methods we could give evidence that $^1O_2$ forms in the Li-O$_2$ battery during discharge and from the onset of charge and that it can account for a major fraction of the side products formed. Hence, $^1O_2$ arises as perhaps the biggest obstacle for cycling of the Li-O$_2$ cell by reversible formation/decomposition of Li$_2$O$_2$. Presence of trace water, which was already known to increase side reactions, acts at least in part by raising the amount of $^1O_2$ generated. We show that $^1O_2$ traps and quenchers can effectively reduce the side reactions on discharge and charge. The level of $^1O_2$ abundance makes traps less likely to be effective for long term cycling since they will be consumed rapidly. Physical quenchers are preferred since they are not consumed. Future work should therefore focus on finding quenchers that are entirely compatible with the cell environment, with the electrochemical potential window, compatibility, and stability against superoxide and peroxide being the most prominent requirements. Equally it needs to be compatible with anodes such as possibly protected Li metal. Alkaline superoxides in the cycling mechanism suggest that the Na-O$_2$ and K-O$_2$ systems would merit investigating whether $^1O_2$ is involved.

**Methods**

**Materials.** Ethylene glycol dimethyl ether (DME, >99.0%), 9,10-dimethylanthracene (DMA, >98.0 %) and 9,10-diphenylanthracene (DPA, >98.%) were purchased from TCI Europe. Tetraethylene glycol dimethyl ether (TEGDME, ≥99%), $d_3$-acetonitrile (≥99.8atom-%), $LiClO_4$ (battery grade, dry, 99.99%), 1,4-diazabicyclo[2.2.2]octane (DABCO, ≥99%) and $H_2O$ (HPLC grade) were purchased from Sigma-Aldrich. APCI/APPI tuning mix was purchased from Agilent Technologies. Formic acid was bought from Fluka Analytical (puriss. p.a. ~98%). Acetonitrile (HiPerSolv Prolabo) was purchased from VWR Chemicals. High purity oxygen ($O_2$ 3.5, >99.95 vol%), high purity Ar (Ar 5.0, >99.999 vol%) and a mixture of Ar 6.0 and $O_2$ 5.5 (Ar ~5 vol%) were purchased from Messer Austria. Moisture content of the solvents and electrolytes was measured by Karl Fischer titration using a TitroLine KF trace (Schott). Solvents were purified by distillation and further dried over activated molecular sieves. $LiClO_4$ was dried under vacuum for 24 h at 160 °C. All chemicals were used without any further purification, except for DABCO, which was purified by recrystallization from absolute diethyl ether. The sensitizer palladium(II) *meso*-tetra(4-fluorophenyl)tetrabenzoporphyrin was synthesized according a previously reported procedure[51]. $Li_2O_2$ was synthesized according to a previously reported procedure[25].

***Operando* and *ex-situ* electrochemical methods and analysis.** Carbon cathodes were fabricated by first making a slurry of Super P carbon (TIMCAL) with PTFE binder in the ratio 9:1 (m/m) using isopropanol. The slurry was then coated onto a stainless steel mesh current collector. The electrodes were vacuum dried at 200 °C for 24 h and then transferred to an Ar filled glove box without exposure to air. The glass fibre separators were washed with ethanol and dried overnight at 200 °C under vacuum prior to use. The $LiFePO_4$ counter electrodes were made by mixing partially delithiated active material with Super P and PTFE in the ratio 8:1:1(m/m/m). The electrodes were vacuum dried at 200 °C for 24 h. The counter electrodes had three-fold the expected capacity of the positive electrode. The electrochemical cells used to investigate cycling were based on a Swagelok design. Typical working electrodes had a carbon mass loading of 1 mg and the cells were assembled with 70 μL electrolyte.

Electrochemical tests were run on either a SP-300 (BioLogic, France) or BT-2000 (Arbin Instruments) potentiostat/galvanostat. Cyclic voltammograms were recorded in a three-electrode arrangement with glassy carbon disc working electrode (BAS Inc.), a Ag wire pseudo-reference and a Pt wire counter electrode inside a glass cell with PTFE lid. The cells were run inside an Ar filled glovebox and purged with high-purity Ar or $O_2$. The redox system $Fc/Fc^+$ was used to reference the measured data *vs.* the $Li/Li^+$ scale.

UV-Vis absorption spectra were recorded on a UV-Vis spectrophotometer Cary 50 (Varian). The molar absorption coefficient of DMA was determined as an average of three independent measurements. Photo-chemical generation of $^1O_2$ was done by *in-situ* photogeneration with the sensitizer palladium(II) *meso*-tetra(4-fluorophenyl)tetrabenzoporphyrin[51]. The sensitizer in the $O_2$ saturated solution was irradiated with a red LED light source (643 nm, 7 W).

Fluorescence Measurements were recorded on a Fluorolog 3 fluorescence spectrometer (Horiba) equipped with a NIR-sensitive photomultiplier R2658 (300 -1050 nm) from Hamamatsu. The *operando* fluorescence measurements were performed in the front face mode in kinetic acquisition mode with 0.1 s excitation every 10 s to minimize photobleaching of the DMA. The fluorophore concentration was adjusted to attain an absorbance of ~0.2 to avoid inner filter effects and to achieve good correlation between the observed fluorescence intensity (proportional to the amount of the absorbed light) and absorption (proportional to the concentration) of the $^1O_2$ trap. DMA was excited at 378 nm and the emission was detected at 425 nm. The cell for *operando* fluorescence was a 1 cm absorption high precision quartz cell (Hellma Analytics) with a purpose made gas-tight PTFE-lid. The working electrode was composed of a Super P/PTFE mixture that was pasted onto a Ti grid. The electrode prefilled with chemically synthesized $Li_2O_2$ was made by mixing the dried electrode material with $Li_2O_2$ and pressing the mixture onto the Ti grid. The reference and counter electrodes were partly delithiated $LiFePO_4$ pasted onto Al grids. The assembling was performed in an Ar filled glovebox. The cell contained a magnetic stirrer bar, was filled with electrolyte, streamed with $O_2$, further connected with a pure $O_2$ reservoir and hermetically sealed before placing it into the spectrometer. During the measurement the electrolyte was stirred to ensure $O_2$ saturation and uniform DMA concentration. The DMA concentration of $1.6\times10^{-5}$ M for the operando fluorescence was chosen to optimize the sensitivity of the method. At an absorbance of A = 0.2 (measurement conditions), 37% of the excitation light is absorbed by the chromophore (= $1-10^{-A}$). In a hypothetical example, reaction of 10% of DMA with $^1O_2$ will decrease absorbance by 10%, i.e. from 0.2 to 0.18. Thus, after the reaction 34% of the excitation light will be absorbed by the chromophore. Since the fluorescence intensity is proportional to the amount of the absorbed light, the decrease of fluorescence intensity will be (37-34)/37*100 = 8%. Analogous calculation with 10 times the DMA concentration (A = 2) results in 99% of excitation light absorbed before bleaching and 98.4% of the excitation light absorbed after bleaching. Thus, the decrease of fluorescence intensity would be (99-98.4)/99*100 = 0.6% which is much lower than for the comparably low concentration of the trap. Therefore, a relatively low concentration of DMA is essential for the best sensitivity of *operando* fluorescence.

*Operando* NIR spectroscopy to detect the emission of singlet oxygen was performed using a germanium detector (model 261, UDT Instruments, Gamma Scientific Company, USA). It was cooled to -30 °C using a Peltier cooling unit. A longpass-filter with a cut-on wavelength of 1200 nm and a shortpass-filter with a cut-off wavelength of 1350 nm (Edmund optics) were placed directly in front of the sensor. The cell for *operando* NIR spectroscopy was a 1 mm absorption high precision quartz cell (Hellma Analytics) with a purpose made gas-tight PTFE-lid. The working electrode was an Au-grid electrode (ALS). The reference and counter electrodes were partly delithiated $LiFePO_4$ attached to an Al-grid. The cell was placed directly in front of the filters followed by an Au-mirror. The optical set-up containing the measurement cell was located in a blackbox to avoid ingress of stray light. The detector signal was amplified by a photodiode amplifier PDA-750 (Tetrahertz Technologies) and the signal recorded on the potentiostat which controlled the cell.

The *operando* electrochemical mass spectrometry setup was built in-house and is similar to the one described previously[52,53]. It consisted of a commercial quadrupole mass spectrometer (Balzers) with a turbomolecular pump (Pfeiffer) that is backed by a membrane pump and leak inlet which samples from the purge gas stream. The electrochemical cell was based on a three-electrode Swagelok design. The setup was calibrated for different gases (Ar, $O_2$, $CO_2$, $H_2$, $N_2$ and $H_2O$) using calibration mixtures in steps over the anticipated concentration ranges to capture nonlinearity and cross-sensitivity. During measurements either a gas mixture consisting of 95% $O_2$ and 5% Ar or pure Ar was used. All calibrations and quantifications were performed using in-house software. The purge gas system consisted of a digital mass flow controller (Bronkhorst) and stainless steel tubing. The procedure for the carbonate / carboxylate analysis was as described earlier[25].

High-performance liquid chromatography coupled with mass spectrometry (HPLC-MS) was used for determining the degree of the DMA to DMA-$O_2$ conversion. The sample handling was performed inside an Ar filled glovebox. The electrolyte was extracted from the cell using DME that was then removed by evaporation at room temperature. The residue was dissolved in 50 μL DME and a volume of 1 μL was injected into the HPLC. The HPLC instrument was a 1200 Series (Agilent Technology, USA) with a multiple wavelength UV-Vis detector (Agilent Technology G1365C MWD SL) coupled to a mass spectrometer using atmospheric pressure chemical ionization (APCI) as ionisation method (Agilent Technologies 6120 Quadrupole LC/MS, Santa Clara, USA). The samples were analysed by a reversed-phase Poroshell column (120 EC-C8, 3.0 mm × 100 mm, Ø 2.7 μm, Agilent Technology, USA) using a gradient system of acetonitrile (solvent B) and water containing 0.01% formic acid (solvent A). A pre-column (UHPLC 3PK, Poroshell 120 EC-C8 3.0 × 5 mm 2.7 μm, Agilent Technology, USA) was connected before the reversed phase column. The elution started with 50% solvent B and was then increased to 100% solvent B within 5 min at a flow rate of 0.7 mL/min. The column was held at 15 °C throughout the measurements. The eluent was monitored via an UV-Vis detector at the wavelengths of 258 nm and 374 nm. The MS signal was recorded starting after 2 min in a mass range of 100 - 450 m/z using the APCI in the positive ion mode. The MS signal was used to identify the retention times for DMA and DMA-$O_2$. The extent of the transformation of 9,10-dimetylanthracene (DMA) to 9,10-dimethylanthracene-endoperoxide (DMA-$O_2$) was determined from the absorbance at 258 nm and the molar absorption coefficients $\varepsilon_{DMA,258nm}$ and $\varepsilon_{DMA-O_2,258nm}$. The latter was determined from DMA-$O_2$, which was obtained by conversion of DMA with photogenerated $^1O_2$.

**Data availability.** The data that support the plots within this paper and other findings of this study are available from the corresponding author upon reasonable request.

**Acknowledgements**

S.A.F. is indebted to the European Research Council (ERC) under the European Union's Horizon 2020 research and innovation programme (grant agreement No 636069). We further gratefully acknowledge funding from the Austrian Federal Ministry of Economy, Family and Youth and the Austrian National Foundation for Research, Technology and Development and initial funding from the Austrian Science Fund (FWF, Project No P26870-N19). The authors thank R. Saf for help with the MS, R. Breinbauer for discussions about the reaction mechanism, S. Landgraf for help with the NIR measurement, and J. Schlegl for manufacturing instrumentation for the methods used.

**Author Contributions**

N.M. performed the main part of the experiments and analysed the results. B.S., S.G. and C.L. performed cell cycling, MS, and NMR experiments. G.A.S. did HPLC analysis. S.A.F., D.K., C.S., O.F., and M.L. discussed the reaction mechanisms. S.M.B. supervised the optical experiments. S.A.F. conceived and directed the research, set up and performed experiments, analysed the results and wrote the manuscript with help of the other authors. All authors contributed to the discussion and interpretation of the results.

**Competing Financial Interests**

The authors declare no competing financial interests